\documentclass{webofc}

\usepackage[varg]{txfonts}   
\usepackage{hyperref}
\usepackage{url}
\hypersetup{colorlinks=true,citecolor=blue,urlcolor=blue,linkcolor=blue}

\newcommand{\bc}{\begin{center}}
\newcommand{\ec}{\end{center}}
\newcommand{\be}{\begin{equation}}
\newcommand{\ee}{\end{equation}}
\newcommand{\bea}{\begin{eqnarray}}
\newcommand{\eea}{\end{eqnarray}}
\newcommand{\ba}{\begin{array}}
\newcommand{\ea}{\end{array}}
\newcommand{\lb}{\label}
\newcommand{\rf}{\ref}
\newcommand{\bfg}{\begin{figure}[htbp]}
\newcommand{\efg}{\end{figure}}

\begin{document}
\title{Scattering of mesons and emergence of tetraquarks \protect \\
       in two-dimensional QCD}
%
%

\author{\firstname{Hagop} \lastname{Sazdjian}\inst{1}
\fnsep\thanks{\email{hagop.sazdjian@ijclab.in2p3.fr}}
}

\institute{Universit\'e Paris-Saclay, CNRS/IN2P3, IJCLab, 91405
Orsay, France}

\abstract{%
Scattering of two mesons is considered in the framework of
two-dimensional QCD in the large-$N_c^{}$ limit with four different
quark flavors. The scattering takes place through two coupled channels,
corresponding to direct and quark-exchange processes, which are of
order $O(N_c^{-2})$ and $O(N_c^{-1})$, respectively. Finiteness of
the scattering amplitudes to order $N_c^{-2}$ is pointed out.
The theory reduces, at low energies, to an effective theory of mesons,
interacting by a quark-exchange process, by means of a contact term.
The unitarization of the scattering amplitudes leads to the emergence
of a tetraquark bound state, located very close to the lowest
two-meson elastic threshold.
}
\maketitle
\section{Introduction} \lb{s1}

Two-dimensional QCD, in the large-$N_c^{}$ limit, first introduced
and analyzed by 't Hooft
\cite{'tHooft:1973jz,'tHooft:1974hx,Callan:1975ps,Einhorn:1976uz},
is an efficient tool for probing many properties of hadrons related
to the confinement of quarks. In two spacetime dimensions,
confinement is a natural property of QCD, due to the strong
infrared behavior of the gluon propagator. In the large-$N_c^{}$ limit,
inelasticity and pair creation effects are damped, in which case the
confinement properties of the theory become more salient. In
noncovariant gauges (light-cone, axial), gluon self-interactions and
ghosts disappear. Finally, in the many-body case, the interactions
between hadronic clusters are free of long-range van der Waals forces.
\par
We investigate, in that framework, the properties of the theory in the
four-body sector, made of two quarks and two antiquarks, each with
different flavor, and search for the possible existence of tetraquark
bound states. The investigation is concentrated on the properties of
two-meson scattering amplitudes. Reviews on the large-$N_c$ limit
can be found in \cite{Witten:1979kh,Coleman:1985rnk} and more
specifically in relation with the tetraquark problem in
\cite{Lucha:2021mwx}.
\par

\section{Quark-antiquark sector} \lb{s2}

The quark fields are considered in the fundamental representation
of the color-gauge group $SU(N_c^{})$.
The light-cone coordinates and momenta are defined as
$x^{\pm}=\frac{1}{\sqrt{2}}(x^0\pm x^1)$,
$p_{\pm}=\frac{1}{\sqrt{2}}(p_0\pm p_1)$. In the light-cone gauge,
the gluon field satisfies $A_-=0$. Because of the particular
structure of the quark-gluon vertex in the light-cone gauge, where
only the Dirac matrix $\gamma_-^{}$ appears, the Dirac matrices can
be factorized from the $N_c^{}$-leading expressions of all dynamical
quantities.
\par
In two dimensions, the coupling constant $g$ of the theory has the
dimension of mass. One may introduce the string tension $\sigma$
according to the definition
\be \lb{e1}
\sigma\equiv \frac{g^2N_c^{}}{4}.
\ee
($g$ has been defined with the one-index notation of the gluon
fields.) In the large-$N_c^{}$ limit, the string tension is assumed
to remain finite, which means that the coupling constant squared
should decrease like $1/N_c$.
\par
The gluon propagator contributes with its $++$ component:
\be \lb{e2}
D(q)\equiv D_{++}(q)=\frac{i}{q_-^2}.
\ee
In two dimensions, this propagator is infrared singular. To avoid
infrared divergences in intermediate calculations, we introduce a
small mass-like parameter $\lambda$ (behaving like the $-$ component
of a momentum) in the propagator, playing the role
of an infrared cutoff; to simplify notations, $\lambda$
will be most often replaced by an inverse quantity, $\Lambda$: 
\be \lb{e3}
\frac{i}{q_-^2}\longrightarrow \frac{i}{q_-^2+\lambda^2},\ \ \ \ \ \
\Lambda\equiv \frac{\sigma}{\lambda}.
\ee
One expects that physical quantities will remain finite when the
limits $\lambda\rightarrow 0$ or $\Lambda\rightarrow\infty$ are taken.
\par
The quark propagator, after taking into account self-energy
radiative corrections coming from the large-$N_c^{}$ leading part,
represented by the one-gluon contribution in its integral equation,
is
\be \lb{e4}
S(p)=\frac{ip_-^{}}{2p_+^{}p_-^{}-|p_-^{}|\Lambda-m^{\prime 2}
+i\epsilon},\ \ \
\ \ \ m^{\prime 2}=m^2-\frac{2\sigma}{\pi}.
\ee
One notices that in the limit $\Lambda\rightarrow\infty$, the
renormalized mass of the quark goes to infinity, indicating
that the quark ceases to be an observable free particle with a finite
mass.
\par
For quark-antiquark systems in the color-singlet sector, one can
write down the corresponding Bethe--Salpeter equation, where the
kernel is given, at leading order of $N_c^{}$, by the one-gluon
exchange diagram. Designating by $r$ the total momentum and by $p$
the quark momentum, one introduces the bound state wave function
$\phi(r,p)$. Noticing that the interaction in the light-cone gauge
is lightlike instantaneous, one further introduces the instantaneous
wave function $\varphi$ defined as
\be \lb{e5}
\varphi(r,p_-^{})=\int\frac{dp_+^{}}{2\pi}\phi(r,p).
\ee
$\varphi$ satisfies spectral conditions that imply that the momentum
component $p_-^{}$ should be positive and bounded by $r_-^{}$, which
itself is positive. Defining then the reduced variable $x$ as
\be \lb{e6}
x=\frac{p_-^{}}{r_-^{}},\ \ \ \ \ \ 0\le x\le 1,
\ee
the equation satisfied by $\varphi$ becomes
\be \lb{e7}
\Big[\,r^2-\frac{m_1^2}{x}-\frac{m_2^2}{(1-x)}\,\Big]\,\varphi(x)=
-\big(\frac{2\sigma}{\pi}\big)\int_0^1dy
\frac{(\varphi(y)-\varphi(x))}{(y-x)^2},
\ee
where $m_1^{}$ and $m_2^{}$ are the quark and antiquark masses,
respectively. This is the 't Hooft equation \cite{'tHooft:1974hx}.
One notices that the infrared cutoff has disappeared, related to
the fact that the eigenvalues $r^2$, which are to be determined,
correspond to the physical meson masses.
\par
The spectrum of the eigenfunctions and eigenvalues corresponds to
an infinite tower of meson states, characterized by a discrete quantum
number $n$, with a linear increase with $n$ of the higher masses
squared, a feature that is typical of a confinement regime.
\par
The quark-antiquark scattering amplitude $\mathcal{T}$ has been
calculated in \cite{Callan:1975ps}. Its expression is
\be \lb{e8}
\mathcal{T}(r,x,x')=-\frac{2\sigma}{N_c^{}}\frac{1}{r_-^2(x-x')^2}
+\frac{1}{N_c^{}}\sum_n\frac{\widetilde \phi_n^{}(r,x)
\widetilde \phi_n^*(r,x')}{(r^2-r_n^2)},
\ee
where $\widetilde \phi_n^{}$ is obtained from $\phi_n^{}$ after
factorizing in the latter the free quark-antiquark Green's function.
One notices that apart from the one-gluon exchange term (the first
term), which cannot be absorbed into products of wave functions,
the scattering amplitude is entirely given by the meson wave function
contributions. The scattering amplitude $\mathcal{T}$ is cutoff
dependent through the complete wave functions $\phi$ or
$\widetilde{\phi}$, which, contrary to the instantaneous wave function
$\varphi$ [Eq. (\rf{e5}], are cutoff dependent. Therefore,
$\mathcal{T}$ is not a directly observable quantity. Furthermore, it
explicitly contains the long-range one-gluon exchange interaction.
In this connection, one of the basic requirements from the theory is
that in observable processes, like meson-meson scattering, the cutoff
dependences, as well as the long-range contributions, disappear from
the final expressions.  
\par

\section{Two-quark--two-antiquark systems} \lb{s3}

We now consider systems made of two quarks and two antiquarks.
The quark flavors are assumed different from each other; this
case avoids mixing problems with ordinary meson states. The
quarks are denoted with labels $1$ and $3$, with masses $m_1^{}$ and
$m_3^{}$, and the antiquarks with labels $\bar 2$ and $\bar 4$, with
masses $m_2^{}$ and $m_4^{}$. The Green's function, corresponding
to the four ingoing and outgoing particles, is schematically
represented in Fig. \rf{f1}.
\bfg 
\bc
\includegraphics[scale=0.6]{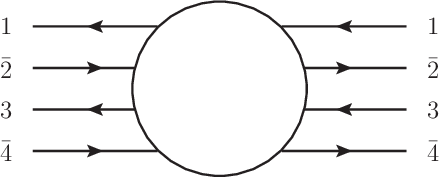}
\caption{Four-body Green's function, with two quarks, $1$ and $3$,
and two antiquarks, $\bar 2$ and $\bar 4$.}
\lb{f1} 
\ec
\efg
\par
We are interested in the color-singlet sectors of the incoming and
outgoing particles in the Green's function, describing scattering
processes of mesons. We have two sets of mesonic clusters:
$(1\bar 2)(3\bar 4)$ and $(1\bar 4)(3\bar 2)$.
One can then distinguish four different channels in the scattering
process, two of which are called ``direct'', the ingoing and outgoing
clusters being the same, and the two others ``recombination'', the
outgoing clusters having undergone a quark exchange. The four
different processes are:
\bea 
\lb{e9}
& &(1\bar 2)+(3\bar 4)\ \longrightarrow (1\bar 2)+(3\bar 4),\ \ \ \
\mathrm{direct\ channel}\ 1\ \ \ (D1),\\
\lb{e10}
& &(1\bar 2)+(3\bar 4)\ \longrightarrow (1\bar 4)+(3\bar 2),\ \ \ \
\mathrm{recombination\ channel}\ 1\ \ \ (R1),\\
\lb{e11}
& &(1\bar 4)+(3\bar 2)\ \longrightarrow (1\bar 4)+(3\bar 2),\ \ \ \
\mathrm{direct\ channel}\ 2\ \ \ (D2),\\
\lb{e12}
& &(1\bar 4)+(3\bar 2)\ \longrightarrow (1\bar 2)+(3\bar 4),\ \ \ \
\mathrm{recombination\ channel}\ 2\ \ \ (R2).
\eea
\par
Integral equations for the Green's functions in the color-singlet
sectors at leading orders of $N_c^{}$ can be obtained following the
method used in the quark-antiquark case. Their detailed derivation
is left to a separate work. We will be content by indicating here
some of the qualitative aspects that characterize them. Similar
derivations and related technical problems can be found in 
\cite{Heupel:2012ua,Eichmann:2015cra,Eichmann:2020oqt,
Kvinikhidze:2021kzu}.
\par
The integral equations relative to the four channels 
decouple into two subsets of equations, the first concerning
channel 1 and the second, channel 2. Inside each subset, one has
two coupled integral equations relative to the direct and
recombination processes. Actually, one passes from one subset to
the other by appropriate permutations of indices. It is therefore
sufficient to concentrate on one subset, channel 1, say.
\par
Specifically, one finds the following large-$N_c^{}$ behaviors
for the scattering amplitudes:
\bea
\lb{e13}
\mathcal{T}_{R1}&=&O(N_c^{-1})+O(N_c^{-3}),\\
\lb{e14}
\mathcal{T}_{D1}&=&O(N_c^{-2})+O(N_c^{-4}).
\eea
The on-mass shell scattering amplitudes are obtained by projecting
the off-mass shell scattering amplitudes on the ingoing and outgoing
wave functions.
\par

\section{Finiteness of the scattering amplitudes to order
\boldmath{$1/N_{\lowercase{c}}^2$}} \lb{s4}

An important test for the theory is the verification that the
meson-meson scattering amplitudes are finite quantities in the
limit $\Lambda\rightarrow \infty$. The infrared cutoff is present
in the quark propagators, the gluon propagator and the external wave
functions. Another test is the verification that the meson-meson
scattering amplitudes are free of long-range van der Waals type
forces. 
\par
The calculations take into account the following features.
Each quark propagator carries a damping factor $\Lambda^{-1}$.
The gluon propagator, upon integration, produces a diverging
factor $\Lambda$. Each external meson wave function carries a
diverging power $\Lambda$. Each quark loop integration with
respect to the $+$ component of the loop momentum removes one
power from the damping factors. Previous calculations of this
type have been presented in \cite{Callan:1975ps}.
\par
Generally, individual diagrams are linearly divergent in $\Lambda$.
One therefore has to consider combinations of several diagrams
having similar structures.
\par
The recombination channel process receives, at order $N_c^{-1}$,
contributions from three diagrams, one of them being free of gluon
propagators and the two others containing one gluon propagator.
It turns out that the sum of the three diagrams is finite in
the limit $\Lambda\rightarrow\infty$. The finite part has the
structure of an effective four-meson contact term with smooth
momentum dependence, excluding any presence of long-range forces.
The effective coupling constant of the contact term is provided by
an overlapping integral of the four external meson wave functions.
\par
Since the next-to-leading term in the recombination channel,
generated by the leading terms in the corresponding Green's function,
is of order $N_c^{-3}$, one should check whether nonleading terms
of the $1/N_c^{}$ expansion might produce divergences at order
$1/N_c^2$. These arise from quark-loop corrections to the gluon
propagator, from vertex corrections and from crossed diagrams.
\par
The quark-loop correction can be calculated using the propagator
(\rf{e4}). The result is a modification of the gluon
propagator (\rf{e2})-(\rf{e3}) to the form \cite{Callan:1975ps}
\be \lb{e15}
D(q)=\frac{i}{q_-^2+\lambda^2+\lambda |q_-^{}|N_f^{}/
(\pi N_c)},
\ee
where $N_f^{}$ is the number of quark flavors. The net effect of
this change is a renormalization of the cutoff parameter $\Lambda$
in the following way:
\be \lb{e16}
\Lambda\longrightarrow\Lambda\,\Big(1-\frac{N_f^{}}{\pi^2N_c^{}}\Big).
\ee
The same change of $\Lambda$ also occurs in the quark propagator
(\rf{e4}), $m^{\prime 2}$ remaining unchanged. Therefore, the
cancellation mechanism between gluon and quark propagator
divergences remains unaffected.
\par
Vertex corrections and gluon crossed diagrams produce generally
leading corrections of order $1/N_c^{}$. However, in the present
case, the latter vanish due to the spectral conditions that the
meson wave functions should satisfy [Eqs. (\rf{e5}) and (\rf{e6})].
\par
Therefore, the finiteness of the recombination channel scattering
amplitude remains true also at order $N_c^{-2}$.
\par
The direct channel process receives, at order $N_c^{-2}$,
contributions from 65 diagrams, containing up to four gluon
propagators. These diagrams can be be grouped, according to their
structure, into five categories, which can be studied independently
from each other. The net result is that the divergent parts
cancel out in each category and one remains with a finite part
which is nothing but the generalized two-meson unitarity diagram
generated by the finite part of the recombination process.
(The two-meson loops are actually infinite in number, involving
all mesons of the quark-antiquark spectra, [Eq. (\rf{e8})].)
\par
The above results establish the finiteness of the two-meson
scattering amplitudes to order $N_c^{-2}$ and provide the
equivalent low-energy effective meson theory, which reduces to
the recombination process contact term and the generation from it
of unitarity diagrams.
\par

\section{Unitarized scattering amplitudes} \lb{s5}

The extension of the above study to higher orders than $N_c^{-2}$
becomes out of reach, because of the increasing number
of diagrams and also because of the appearance of new types of
diagrams. For the last category, we might observe that even if they
produce new types of effective interactions, they could not compete,
at large $N_c^{}$, with those found up to order $N_c^{-2}$. It is
therefore not unreasonable to ignore, for a first study, such
possible new effects.
\par
Concerning the higher-order contributions generated by the
leading-order diagrams through the Green's functions, we observe
that the results obtained above indicated the tendancy of the theory
to unitarize the effective contact interaction generated by the
recombination process. Assuming that the resulting effective theory
satisfies the unitarity property, one is entitled to complete the
results obtained up to order $N_c^{-2}$ by unitarizing the
corresponding scattering amplitudes.
\par
We undertake that operation using two approximations. First,
we keep in the series of intermediate meson states that contribute
to the unitarity loop diagrams only the ground state mesons,
neglecting their radial excitations. The latter contribute with
dampted mass factors and with smaller four-meson couplings, due
to the smallness of the overlapping integrals between ground state
and radial wave functions. Second, we neglect the momentum
dependence of the effective four-meson coupling of the ground-state
mesons, by treating it as a constant. With these approximations,
the relevant equations are reduced to coupled four-channel
equations [Eqs. (\rf{e9})-(\rf{e12})], that could be dealt with in
matrix form.
\par
Designating by $K_R^{}$ the real coupling constant
of the recombination channel (unitarity requiring the equality
$K_{R1}^{}=K_{R2}^{}$), the scattering amplitude and the kernel of
its integral equation take the following matrix forms:
\be \lb{e17}
\mathcal{T}=\left( \ba{cc}
\mathcal{T}_{D1}^{} & \mathcal{T}_{R2}^{} \\
\mathcal{T}_{R1}^{} & \mathcal{T}_{D2}^{}
\ea \right),\ \ \ \ 
K=\left( \ba{cc}
0 & {K}_{R}^{} \\
{K}_{R}^{} & 0
\ea \right),\ \ \ \ 
J=\left( \ba{cc}
J_{1\bar 2,3\bar 4}^{} & 0 \\
0 & J_{1\bar 4,3\bar 2}^{}
\ea \right),
\ee
where we have also introduced the two-meson loop functions
$J_{1\bar 2,3\bar 4}^{}$ and $J_{1\bar 4,3\bar 2}^{}$:
\bea \lb{e18}
& &J_{a\bar b,c\bar d}^{}(s,M_{a\bar{b}}^{},M_{c\bar{d}}^{})
=-\frac{i}{2\sqrt{-\lambda_{a\bar b,c\bar d}^{}}}\Big[1-
\frac{1}{\pi}\arctan\Big(\frac{\sqrt{-\lambda_{a\bar b,c\bar d}^{}}}
{s-(M_{a\bar{b}}^{2}+M_{c\bar{d}}^{2})}\Big)\Big],\nonumber \\    
& &\lambda_{a\bar b,c\bar d}^{}=
\Big(s-(M_{a\bar{b}}^{}+M_{c\bar{d}}^{})^2\Big)
\,\Big(s-(M_{a\bar{b}}^{}-M_{c\bar{d}}^{})^2\Big).
\eea
\par  
When the interaction is represented by a contact term, as is the
case here, the unitarization operation reduces to the iteration of
the interaction by means of the meson loops. One has
\bea \lb{e19}
\mathcal{T}&=&(1-iKJ)^{-1}K \nonumber \\
&=&\frac{1}{\Big(1+K_R^{}J_{1\bar 4,3\bar 2}^{}
K_R^{}J_{1\bar 2,3\bar 4}^{}\Big)}
\left( \ba{cc}
iK_R^{}J_{1\bar 4,3\bar 2}^{}K_R^{} & K_R^{} \\
K_R^{} & iK_R^{}J_{1\bar 2,3\bar 4}^{}K_R^{}
\ea \right),
\eea
from which one may identify, with (\rf{e17}), the different components
of $\mathcal{T}$.
\par

\section{Tetraquarks} \lb{s6}

Tetraquark bound states may occur at the zeros of the denominator of
$\mathcal{T}$. This leads to the equation
\be \lb{e20}
1+K_R^{}J_{1\bar 4,3\bar 2}^{}K_R^{}J_{1\bar 2,3\bar 4}^{}=0.
\ee
This equation has generally one bound state solution, which
lies very close to the lowest elastic two-meson threshold, with
binding energies of the order of 1 MeV or much less, according to
the mass configurations of the mesons.  The smallness of the binding
energy is traced back to the smallness of the coupling constant
$K_R^{}$, which, within a rough estimate, is of the order of
$10\,(2\sigma/\pi)$, with respect to some combinations of the meson
masses. In the case of two heavy quarks, the latter ratio effectively
contributes with a power of 4.  
\par
For the particular example of the configuration $(c\bar{s}b\bar{u})$,
which leads, in the pseudoscalar sector, to the two two-meson clusters
$(D^+B^-)$ and $(D^0\bar{B}_s^0)$, the binding energy comes out of
the order of 0.02 MeV with respect to the $(D^0\bar{B}_s^0)$
elastic threshold. 
\par
Existence of tetraquark bound states with four different quark
flavors, containing $b$ and $c$ quarks, has been observed in Lattice
QCD calculations \cite{Alexandrou:2023cqg,Padmanath:2023rdu,
Mathur:2021gqn}.
\par
Of particular interest is the comparison of the coupling constants
of the tetraquark state to the two meson clusters. Assuming for
definiteness that
$(M_{1\bar 2}^{}+M_{3\bar 4}^{})>(M_{1\bar 4}^{}+M_{3\bar 2}^{})$,
and denoting the tetraquark binding energy by $E_{TB}^{}$ and
$\Delta M=(M_{1\bar 2}^{}+M_{3\bar 4}^{})-
(M_{1\bar 4}^{}+M_{3\bar 2}^{})$, one observes from Eqs. (\rf{e18})
and (\rf{e19}), for nonrelativistic systems, that the
coupling constant squared of the tetraquark to the clusters
$(1\bar 2)(3\bar 4)$ is proportional to $E_{TB}^{-1/2}$,
while that to the clusters $(1\bar 4)(3\bar 2)$ is proportional to
$(\Delta M)^{-1/2}$. The ratio of these quantities for the example
considered above is of the order of 30, which means that the
tetraquark couples more strongly to the heavier two-meson clusters
than to its neighboring ones.
\par
The latter property would suggest at first sight that the tetraquark
is essentially a molecular system made of the heavier cluster
ingredients, here $(1\bar 2)(3\bar 4)$. However, such an
interpretation, neglecting the second set of clusters, would not
explain in a natural way the particular
position of the tetraquark pole in the vicinity of the lighter
clusters. The two two-meson clusters seem to play an equal role.
This situation is due to the quark-exchange mechanism, which is the
basic interaction ingredient of the two-meson clusters, and which
needs the presence of the two distinct two-meson clusters to
operate.
\par

\section{Conclusion} \lb{s7}

Two-dimensional QCD at large $N_c^{}$ allows a consistent treatment
of the confining interactions, with possible control of the infrared
divergences. In the sector of four different flavors, the theory
reduces, at low energies, to an effective theory of mesons, where
the basic interaction is provided by a quark-exchange four-meson
contact term. The unitarization of the scattering amplitude in all
channels of interaction leads to the prediction of the existence of
a tetraquark bound state lying very close to the lightest two-meson
elastic threshold. 
\par
\begin{acknowledgement}
The author thanks W. Lucha and D. Melikhov for stimulating
discussions on the subject. This research has received financial
support from the EU research and innovation programme Horizon 2020,
under Grant agree\-ment No. 824093.
\end{acknowledgement}  

%
%
%

\end{document}